\documentclass[fleqn,usenatbib,revision]{mnras}

\usepackage{graphicx}	
\usepackage{amsmath}	
\usepackage{amssymb}	
\usepackage[dvipsnames]{xcolor}
\usepackage{hyperref}

\usepackage{newtxtext,newtxmath}
\usepackage{ulem}
\usepackage[T1]{fontenc}

\DeclareRobustCommand{\deLima}[3]{#2}
\let\deLimathebibliography\thebibliography
\def\thebibliography{\DeclareRobustCommand{\deLima}[3]{##3}\deLimathebibliography}

\begin{document}



\title[Thin Disks with the Quadrudipole Magnetic Field]{Magnetically Threaded Thin Disks in the Presence of the Quadrupole Magnetic Field}


\author[\c{C}\i{}k\i{}nto\u{g}lu]{
Sercan \c{C}\i{}k\i{}nto\u{g}lu$^{1}$\thanks{E-mail: cikintoglus@itu.edu.tr}
\\
$^{1}$Istanbul Technical University, Faculty of Science and Letters,
Physics Engineering Department, 34469, Istanbul, Turkey}

\date{Accepted XXX. Received YYY; in original form ZZZ}

\pubyear{2023}

\label{firstpage}
\pagerange{\pageref{firstpage}--\pageref{lastpage}}
\maketitle

\begin{abstract}
Neutron stars might have multipole magnetic fields as implied by recent observations
of pulsars.
The presence of the quadrupole field might have an effect on the interaction between the disc and the neutron star
depending on the location of the inner radius of the disc and the strength of the quadrupole field.
For a quadrudipole stellar field, we calculate the toroidal field generated within the disc,
the magnetospheric radius and the torque exerted onto the star. Also, we deduce the effect of
the rotation of the star on the magnetospheric radius which is relevant even for pure dipole magnetic fields.
\end{abstract}


\section{Introduction}
The magnetically threaded disc model is introduced in the pioneering works of
Ghosh \& Lamb \citep{Ghosh77,Ghosh78,Ghosh79a,Ghosh79b} to explain spin evolutions
of the X-ray pulsars.
According to the model, stellar magnetic field lines penetrate the disc over a 
broad region. While the magnetic field lines corotate with the star outside the disc,
they sharply transit to the Keplerian rotation within the disc as they are drifted by the disc flow.
This local angular velocity difference generates a toroidal magnetic field from the polar component of the stellar dipole magnetic field.
On the other hand, some physical processes limit the toroidal magnetic field generation \citep{Ghosh79a,Wang87,Campbell92}.
As a result, a toroidal magnetic field proportional to the polar component of
the stellar dipole magnetic field is formed in steady-state 
where the proportionality factor is a function of the local angular frequency difference
between the star and the fluid of the disc as well as the time-scale of the process which limits the toroidal magnetic field generation.

In this model, the material stresses are balanced by the magnetic stresses
at the so-called \textit{magnetospheric radius} \citep{Ghosh78}. 
The magnetic field dominates
the flow inside this radius and the disc flow deviates from the Keplerian to be aligned with
the stellar magnetic field within a narrow region (boundary layer) \citep{Scharlemann78,Ghosh79a}. 
Depending on the location of the magnetospheric radius relative to 
the corotation radius where the angular frequency of the disc flow equals the star's angular frequency, 
the disc matter is accreted onto the star or expelled \citep{Illarionov75}. 

The star experiences torque because of two factors.
Firstly, the accreted matter also carries angular momentum \citep{Pringle72}.
Secondly, the magnetic stresses that arose due to the interaction of the stellar magnetic field
with the disc, specifically with the toroidal magnetic field generated inside the disc,
exert a torque onto the star \citep{Ghosh78}.

The description of how the stellar field interacts with the disc in Ghosh and Lamb's model 
was later revised by \citet{Kaburaki86,Wang87,Wang96}. The magnetospheric radius
given in these works has the same scaling dependence on the stellar magnetic field and mass-accretion rate, however, 
its numerical prefactor is slightly different from Ghosh and Lamb's model. Also, more moderate suggestions are 
introduced for the magnetospheric radius when it is not inside the corotation radius
either because of the low mass-accretion rate or fast rotation of the star \citep{Rappaport04,Ertan17}.
Despite all these revisions, Ghosh and Lamb's model can still be regarded as a solid model
since it successfully explains many observational features of X-ray pulsars.

In all these works, the magnetic field of the neutron star is considered as a pure dipole magnetic field.
However, higher multipole fields can be formed during the formation of the star \citep{Ardeljan05}
or in the process of mass accretion \citep{Suvorov20}. Such higher multipole fields are
suggested to explain pulse profiles of millisecond pulsars \citep{Krolik91},
X-ray emission properties of millisecond pulsars \citep{Grindlay02,Cheng03},
and, recently, the super-Eddington luminosities of pulsating ultra-luminous X-ray sources \citep{Eksi15,Brice21},
and fast radio bursts from magnetars \citep{Yamasaki22}.
Also, the existence of higher multipole fields 
is favoured by recent observations of millisecond pulsars 
\citep{Bilous19,Miller19,Riley19,Kalapotharakos21} and soft gamma repeaters \citep{Guver11,Lima20}.

The inner disc radius is estimated for pure higher-order multipole fields 
by only generalising the scale of the radial profile of the stellar magnetic field
in some works \citep{Lipunov78,Psaltis99,Long12,Alpar12}. These authors either use the condition of the balance of the pressures or 
the balance of the stresses. The former is suitable for spherical accretion
rather than disc accretion. When the latter condition is used, one has to take into account
the contribution of the multipole fields into the generation of the toroidal magnetic field
within the disc as well as the additional stress terms that arose due to the non-vanishing components
of the stellar magnetic field. None of these are considered in those works.  
Furthermore, in a realistic setting, the higher-order fields should co-exist with the dipole field.

In this work, in the presence of the quadrupole stellar magnetic field in addition to the dipole field (quadrudipole field),
we calculate the steady-state toroidal magnetic field generation within the disc (Section~\ref{sec:bphi}), 
the magnetospheric radius (Section~\ref{sec:rmag}),
and the magnetic torque exerted onto the star as a result of the disc-magnetosphere interaction (Section~\ref{sec:torque})
by following work of \citet{Ghosh78} as well as \citet{Wang87}.

\section{Quadrudipolar Magnetic field}
The components of a quadrudipolar magnetic field of the star is given by
\begin{align}
B^r_*=&\mu\left[\frac{\cos\theta}{r^3}+\frac{f_{Q}}{2}\left(3\cos^2(\theta-i_Q)-1\right)\frac{R}{r^4}\right]\,,\\
B^\theta_*=&\mu\left[\frac{\sin\theta}{2r^3}+f_{Q}\cos(\theta-i_Q)\sin(\theta-i_Q)\frac{R}{r^4}\right]\,.
\end{align}
Here, $\mu$, $R$, $f_Q$, $i_Q$ are, respectively, the magnetic dipole moment, the radius of the star,
the ratio of the strength of the dipole field to the quadrupole field at the star's surface,
and the inclination angle between fields.
We will consider quadrupole field to be aligned with the dipole field, i.e.,~$i_Q=0$ for simplicity throughout the paper.
Since the disc is modelled as a thin disc, all quantities are evaluated around the disc-midplane, 
i.e.,\ $\theta\simeq\pi/2$. Accordingly, the stellar magnetic field
reduces to
\begin{align}
B^r_*\left(\theta\rightarrow\pi/2\right)=&\,-\mu\frac{f_{Q}}{2}\frac{R}{r^4}\,,
\label{eq:Br}\\
B^\theta_*\left(\theta\rightarrow\pi/2\right)=&\,\frac{\mu}{2r^3}\,,
\label{eq:Bz}
\end{align}
around the disc-midplane.
Obviously, in the presence of the quadrupole field, the radial component of the stellar magnetic field does not 
vanish unlike the pure dipole case.

\section{Steady-state Toroidal Magnetic Field}
\label{sec:bphi}
The evolution of the toroidal magnetic field is governed by the induction equation,
\begin{equation}
\frac{\partial \vec{B}}{\partial t}=
\vec{\nabla}\times \left( \vec{v}\times \vec{B}\right)
+\frac{c^2}{4\pi \sigma_{\rm eff}}\nabla^2 \vec{B}\,,
\end{equation}
where $\sigma_{\rm eff}$ is the effective conductivity within the disc.
When the terms on the right hand side of the equation balance
each other, the toroidal magnetic field reaches its steady-state 
value. 

In cylindrical coordinates, the $\phi$-component
of the first term on the right hand side of the equation 
can be written as
\begin{equation}
\frac{\partial}{\partial z}\left(v^\phi B^z - v^z B^\phi\right)
-\frac{\partial}{\partial r}\left(v^r B^\phi - v^\phi B^r\right)
\simeq 
\frac{\partial v^\phi B^z}{\partial z}
+\frac{\partial v^\phi B^r}{\partial r}
\end{equation}
where we use the approximation of $v^\phi\gg v^r\gg v^z$. Also,
we assume the penetrated stellar magnetic field lines are undistorted.
Therefore, we can use expressions \eqref{eq:Br}-\eqref{eq:Bz} for radial and vertical components 
of the magnetic field inside the disc,
\begin{align}
B^r=&\,-\mu\frac{f_{Q}}{2}\frac{R}{r^4}\,,\\
B^z=&\,\frac{\mu}{2r^3}\,.
\end{align}

The vertical variation of $B^z$ can be neglected around the disc-midplane.
On the other hand, the angular frequency sharply transits to the Keplerian angular frequency, 
$\Omega_K=\sqrt{GM/r^3}$, from the corotation, $\Omega$, within a finite depth.
The depth of this transition can be estimated as the half-thickness of the disc, $h$. 
However, a more general estimation would be considering the depth as $d$ such that
$d \leq h$. 
Accordingly, the vertical derivative of the angular velocity can be written as
\begin{equation}
\frac{\partial}{\partial z}\left(v^\phi B^z\right)
\simeq \pm \delta_z(\Omega-\Omega_{\rm K}) B^z\,,
\end{equation}
where $\delta_z\equiv r/d$ is a dimensionless numerical factor.
The above expression takes the positive sign above the disc-midplane
while it takes the negative sign below the disc-midplane.

The radial derivative term can be calculated simply 
by using Keplerian angular velocity profile for $v^\phi$
and expression \eqref{eq:Br} for the radial component of the magnetic field,
\begin{equation}
\frac{\partial v^\phi B^r}{\partial r}
=-\frac{\partial}{\partial r}\left(\sqrt{\frac{GM}{r}}\frac{\mu f_Q R}{2r^4}\right)
=-\frac{9}{2}\Omega_{\rm K}B^r.
\end{equation}

By following works of \citet{Ghosh79a} and \citet{Wang87}, we write the decay term in the induction equation as
\begin{align}
\frac{c^2}{4\pi \sigma_{\rm eff}}\nabla^2 B^\phi
\simeq \frac{B^\phi}{\tau_\phi}\,,
\end{align}
where $\tau_\phi$ is the time-scale of the mechanism that limits the growth of the toroidal magnetic field.
There are a few possible mechanisms introduced, for instance,
reconnection of the opposite polarity toroidal fields across the disc-midplane,
reconnection of the stellar field with the disc, the buoyancy of the toroidal field 
\citep[see][and references therein]{Wang95,Li96}.
If the diffusive decay because of the turbulent motion within the disc \citep{Campbell92,Yi95} is
considered, the time-scale is given as 
\begin{equation}
\tau_\phi=\frac{1}{\alpha\Omega_K}\,,
\end{equation}
\citep{Wang95} where $\alpha$ is a numerical parameter related with the viscosity within the disc \citep{Shakura73}.
Therefore, the steady-state toroidal magnetic field generated within the disc is
\begin{equation}
B^\phi = 
\pm\gamma_\phi \frac{\Omega_{\rm K}-\Omega}{\Omega_{\rm K}} B^z 
+\frac{9B^r}{2\alpha}\,,
\label{eq:bphi}
\end{equation}
where $\gamma_\phi=\delta_z/\alpha$. The factors on the $B^z$ is called as the pitch factor.

\section{Magnetospheric Radius}
\label{sec:rmag}
\begin{figure*}
\center
\includegraphics{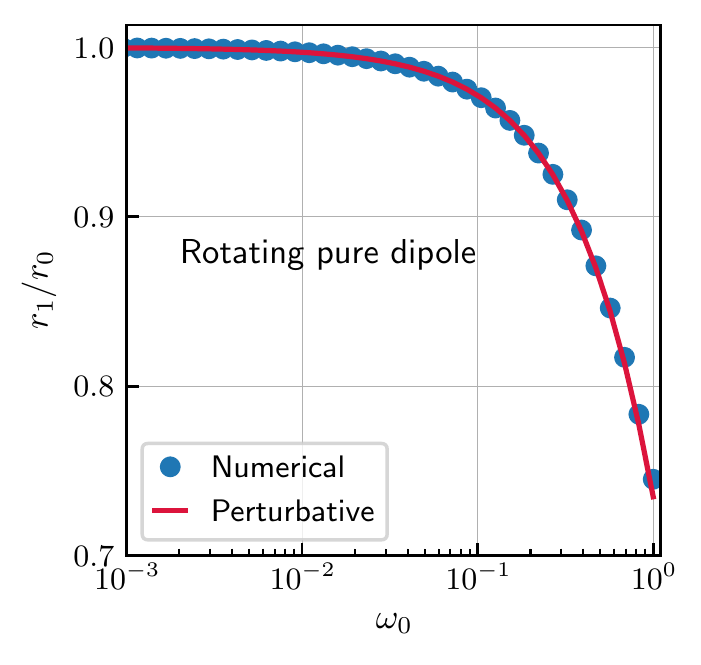}
\includegraphics{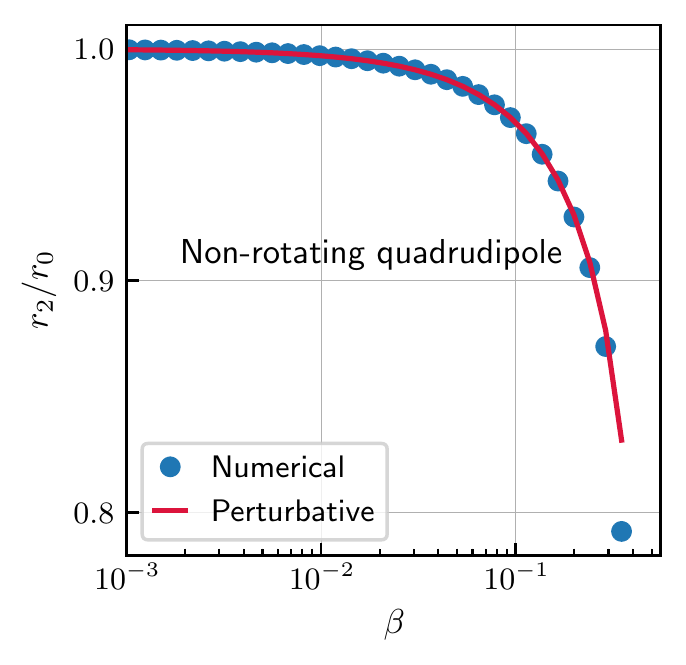}
\caption{The magnetospheric radius of a rotating pure dipole star (left panel) and of a non-rotating quadrudipole star (right panel).}
\label{fig:r12}
\end{figure*}

The magnetospheric radius is defined as where the magnetic field stresses are comparable with the material stresses, hence,
the Keplerian disc flow transforms to be aligned with the magnetic fields inside the magnetospheric radius.
Therefore, the magnetospheric radius can be deduced from the conservation of the angular momentum in the steady-state,
\begin{equation}
\frac{\rho v^r}{r} \frac{\partial}{\partial r}\left(r v^\phi\right) = 
\frac{1}{4\pi}\left[
\frac{\partial}{\partial z}\left(B^z B^\phi\right)
+\frac{1}{r^2}\frac{\partial}{\partial r}\left(r^2 B^r B^\phi\right)
\right],
\label{eq:stresses2}
\end{equation} 
which also corresponds to the condition of the balance of the stresses.
Note that, the radial component of the magnetic field at the disc-midplane 
vanishes for the pure dipole case. However, this term remains in the presence of the quadrupole field.

If expression~\eqref{eq:bphi} is used for the toroidal magnetic field,
and then the vertical average of the equation is taken over the thickness of the disc,
the first term of the right hand side becomes a boundary term,
\begin{align}
\frac{1}{2h}\int_{-h}^{h} \mathrm{d}z\,\frac{\partial}{\partial z}\left(B^z B^\phi\right) 
=&\, 
\frac{1}{2h}\left[(B^z B^\phi)^+ - (B^z B^\phi)^-\right] \notag \\
=&\,\frac{\gamma_\phi}{h} \left(1-\frac{\Omega}{\Omega_{\rm K}}\right) (B^z)^2,
\end{align}
while the second term gives,
\begin{equation}
\frac{1}{2h}\int_{-h}^{h} \frac{1}{r^2}\frac{\partial}{\partial r}  r^2 B^r B^\phi\,\mathrm{d}z \simeq
\frac{1}{r^2}\frac{\partial}{\partial r} \frac{9r^2}{2\alpha}(B^r)^2
=
-\frac{27}{\alpha r}(B^r)^2,
\end{equation}
since the first term in equation~\eqref{eq:bphi} is anti-symmetric and the second term is symmetric across the disc-midplane.
The vertical average of the term on left hand side of equation~\eqref{eq:stresses2} combining with the continuity equation can be written as
\begin{equation}
\frac{\rho v^r}{r} \frac{\partial}{\partial r}\left(r v^\phi\right)
=\frac{\dot{M}}{4\pi r^2 h} \frac{\partial}{\partial r}\left(r^2 \Omega_{\rm K}\right)
=\frac{\dot{M}}{8\pi h} \sqrt{GM} r^{-5/2}\,.
\end{equation}
Therefore, the magnetospheric radius can be deduced from
\begin{equation}
\frac{\dot{M}}{2h} \sqrt{GM} r^{-5/2}=
\frac{\gamma_\phi}{h} \left(1-\frac{\Omega}{\Omega_{\rm K}}\right)  (B^z)^2
-\frac{27}{\alpha r}(B^r)^2\,.
\label{eq:bal_main}
\end{equation}
Accordingly, the presence of the quadrupole term does not contribute to the magnetic stress of $B^z B^\phi$.
However, it yields an additional magnetic stress term, i.e.,~$B^r B^\phi$. Because of its direction,
this term decreases the magnetospheric radius compared with the pure dipole case. Moreover,
a pure quadrupole field cannot hold the radial flow of the disc at the disc-midplane. 

For a non-rotating star with a pure dipole magnetic field, the magnetospheric radius is simply,
\begin{equation}
r_0=\xi\left(\frac{\mu^2}{\sqrt{2GM}\dot{M}}\right)^{2/7}\,,
\label{eq:r0}
\end{equation}
where $\xi=(\gamma_\phi/\sqrt{2})^{2/7}$. This expression is the conventional magnetospheric radius in the case of a pure dipole
stellar field. The prefactor $\xi$ might be defined differently depending on the assumptions and approximations
in the calculations of stresses. Its numerical value is estimated as $0.52$ in the Ghosh and Lamb's model \citep{Ghosh79b} and
it is constrained in the range of $0.35-1.2$ by observations \citep{Psaltis99,Erkut04,DallOsso16}. 
Also, magnetohydrodynamics simulations favour $\xi\sim 0.5$ \citep{Romanova02,Romanova03,Romanova04}.

Note that above expression for the magnetospheric radius is valid only if $\Omega\ll \Omega_K\left(r_{\rm msph}\right)$
even for a pure dipole magnetic field. For taking the dependence of the pitch factor on the star's angular frequency
into account in case of the pure dipole field, one has to solve,
\begin{equation}
\left(\frac{r}{r_0}\right)^{7/2}+\omega_0\left(\frac{r}{r_0}\right)^{3/2}=1\,,
\end{equation}
where the nominal fastness parameter is defined as
\begin{equation}
\omega_0\equiv\left(\frac{r_0}{R_{\rm co}}\right)^{3/2}\,,
\end{equation}
and the corotation radius where the disc flow rotates with the angular frequency of the star, is
\begin{equation}
R_{\rm co}=\left(\frac{GM}{\Omega^2}\right)^{1/3}.
\end{equation}
If the solution is introduced as
\begin{equation}
r_1=r_0\left(1+\zeta_1\right)^m,
\end{equation}
and then both  $\zeta_1$ and $\omega_0$ are considered as small corrections,
the approximate solution can be found as
\begin{equation}
r_1=r_0\left(1-\frac{\omega_0}{7}\right)^2\,,
\label{eq:r1}
\end{equation} 
for a rotating pure dipole star. The comparison of this approximate solution with
the numerical solution is given in the left panel of Fig.~\ref{fig:r12}.
Accordingly, the approximate solution is well-consisted with the numerical solution
and the error is below $2\%$. Note that, the solution is valid only for $r_{\rm msph}<R_{\rm co}$.

\begin{figure*}
\centering
\includegraphics{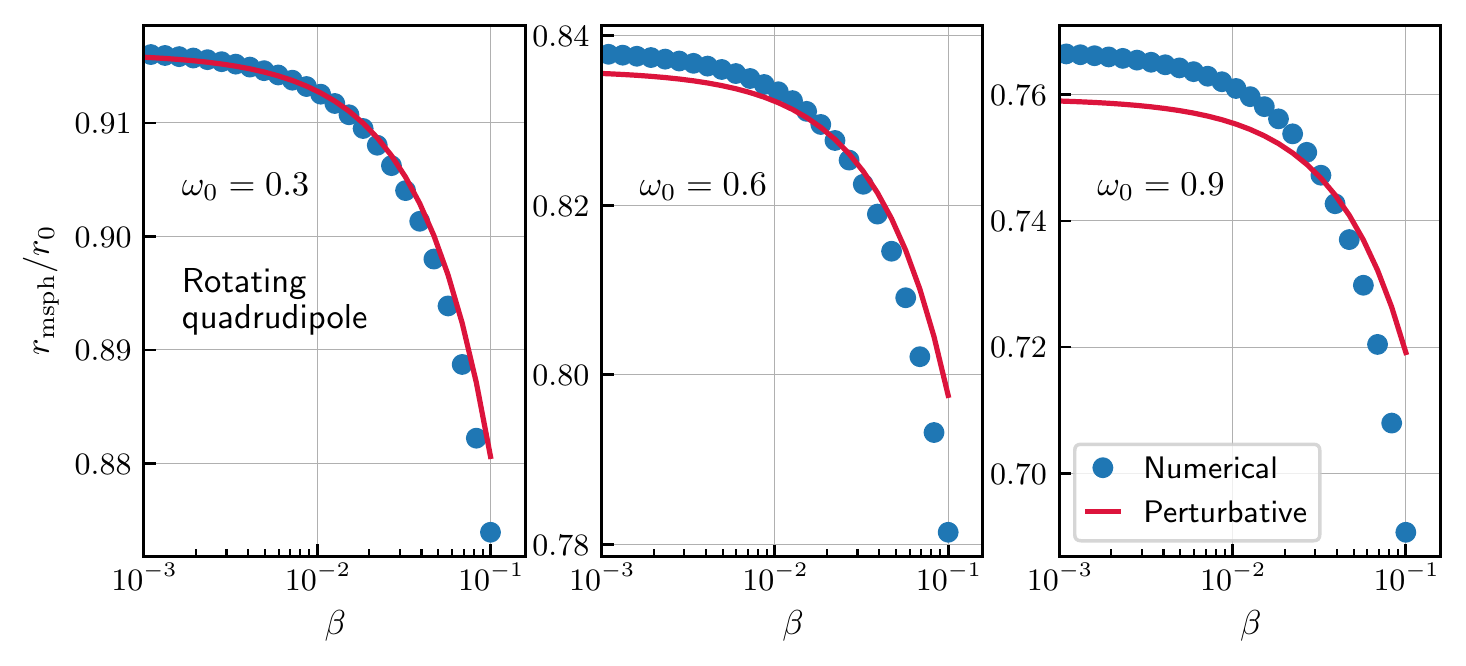}
\caption{The magnetospheric radius of a rotating quadrudipole star for some values $\omega_0$ and $\beta$.}
\label{fig:rmsph}
\end{figure*}

On the other hand, for a non-rotating star with a quadrudipole field,
equation~\eqref{eq:bal_main} becomes
\begin{equation}
\left(\frac{r}{r_0}\right)^{7/2}+\beta\left(\frac{r}{r_0}\right)^{-2}=1\,,
\label{eq:bal_quad}
\end{equation}
where
\begin{equation}
\beta \equiv  \frac{27}{\alpha\sqrt{2}\xi^{7/2}}\frac{h}{r} f_Q^2 \frac{R^2}{r_0^2}\,.
\label{eq:beta2}
\end{equation}
Note that, the ratio of $h/r$ is so-called the thickness parameter and it is a constant
through the disc.
When the second term is dominant in equation~\eqref{eq:bal_main}, it is obvious that 
there is not any real solution. In this case, the disc matter accretes onto the star
through the disc midplane by following the radial magnetic field.
The upper limit for $\beta$ can be deduced from equation~\eqref{eq:bal_quad},
and after a simple analysis, it is found that
the left hand side of equation~\eqref{eq:bal_quad} is always greater than one when $\beta\gtrsim0.35$.
If the magnetospheric radius is introduced as 
\begin{equation}
r_2 = r_0\left(1+\zeta_2\right)^m,
\end{equation}
equation~\eqref{eq:bal_quad} becomes
\begin{equation}
\left(1+\zeta_2\right)^{7m/2}+\beta\left(1+\zeta_2\right)^{-2m}=0\,.
\end{equation}
By considering $\zeta_2$ and $\beta$ much smaller than unity, and expanding the equation up to the second order,
the magnetospheric radius is found as
\begin{align}
r_2 = r_0\left(1-\frac{15}{7}\beta\right)^{2/15},
\label{eq:r2}
\end{align}
for non-rotating quadrudipole field. The difference between the approximate solution
and the numerical solution increases as $\beta$ increases but still the error does not exceed $5\%$ (see the right panel of Fig.~\ref{fig:r12}).

For general case, equation~\eqref{eq:bal_main} can be written as
\begin{align}
\left(\frac{r}{r_0}\right)^{7/2}+\omega_0\left(\frac{r}{r_0}\right)^{3/2}+\beta\left(\frac{r}{r_0}\right)^{-2}=1\,.
\label{eq:bal_gen}
\end{align}
Combining expressions \eqref{eq:r1} and \eqref{eq:r2}, we introduce the magnetospheric radius for the general case as
\begin{align}
r_{\rm msph} = r_0\left(1-\frac{\omega_0}{7}\right)^2\left(1-\frac{15}{7}\beta\right)^{2/15}\left(1-\zeta_3\right)\,,
\end{align}
where $\zeta_3$ is a second order correction arose due to the coupling of $\omega_0$ and $\beta$. 
After inserting this expression into equation~\eqref{eq:bal_gen} and expanding up to second-order,
$\zeta_3$ is found as $(12/49)\omega_0\beta$. Therefore, the magnetospheric radius  
for a rotating star with a quadrudipole magnetic field is
\begin{align}
r_{\rm msph} = r_0\left(1-\frac{\omega_0}{7}\right)^2\left(1-\frac{15}{7}\beta\right)^{2/15}\left(1-\frac{12}{49}\omega_0\beta\right)\,.
\label{eq:rmsph}
\end{align}
We report the magnetospheric radius for various value of $\omega_0$ and $\beta$ in Fig.~\ref{fig:rmsph}.
Accordingly, the difference between the numerical and the approximate solution given in equation~\eqref{eq:rmsph} increases 
as either $\omega_0$ or $\beta$ increases. Still, the error is always less than $5\%$.

\section{Torque Exerted onto the Star}
\label{sec:torque}
The magnetic torque exerted onto the star due to the interaction of the stellar field with the disc is given by
the integral of the magnetic stress over the surface of the disc,
\begin{equation}
N_{\rm mag}=\int_{r_{\mathrm{msph}}}^{\infty} \mathrm{d}r\,\frac{r^2B^\phi B^z}{2}  \left.\right|_{z\rightarrow 0^+}- 
\int_{r_{\mathrm{msph}}}^{\infty} \mathrm{d}r\,\frac{r^2B^\phi B^z}{2}  \left.\right|_{z\rightarrow 0^-}\,.
\end{equation}
Note that, since the second term in the equation~\eqref{eq:bphi} and $B^z$ are symmetric and the first term in the equation~\eqref{eq:bphi} is anti-symmetric across the disc-midplane,
these integrals reduce to
\begin{align}
N_{\rm mag}=&\,\int_{r_{\mathrm{msph}}}^{\infty} \mathrm{d}r\,r^2  \gamma_\phi \frac{\Omega_{\rm K}-\Omega}{\Omega_{\rm K}} (B^z)^2 
\notag \\
=&\,\gamma_\phi (B^z)^2 r_{\rm msph}^3 \frac{1-2\omega_*}{3}
\notag \\
=&\,\dot{M}\sqrt{GMr_{\rm msph}}\left[1+\beta\left(\frac{r_0}{r_{\rm msph}}\right)^{11/2}\right]
 \frac{1-2\omega_*}{6(1-\omega_*)}
\notag \\
\simeq&\,\dot{M}\sqrt{GMr_{\rm msph}}\left(1+\beta\right)
 \frac{1-2\omega_*}{6(1-\omega_*)}\,,
\end{align}
where the fastness parameter is defined as
\begin{equation}
\omega_* = \frac{\Omega}{\Omega_K\left(r_{\rm msph}\right)}\,.
\end{equation}
Therefore, the total torque can be written as
\begin{equation}
N_{\rm tot}=\dot{M}\sqrt{GMr_{\rm msph}}+N_{\rm mag}
= \dot{M}\sqrt{GMr_{\rm msph}}\,n\left(\omega_*\right)
\end{equation}
where the dimensionless torque is
\begin{equation}
n\left(\omega_*\right)=\frac{7+\beta-(8-2\beta)\omega_*}{6\left(1-\omega_*\right)}\,.
\label{eq:ntorque}
\end{equation}
The critical fastness parameter is defined as where the dimensionless torque, hence,
the total torque becomes zero. Accordingly, it is deduced as
\begin{equation}
\omega_c=\frac{7+\beta}{8-2\beta}
\simeq \frac{7}{8}+\frac{11}{32}\beta\,,
\end{equation}
from equation~\eqref{eq:ntorque}.
Considering the maximum value of $\beta$ being $0.35$, 
the presence of the quadrupole field shifts the critical fastness parameter
from $0.875$ to $0.995$.

\begin{figure}
\center
\includegraphics{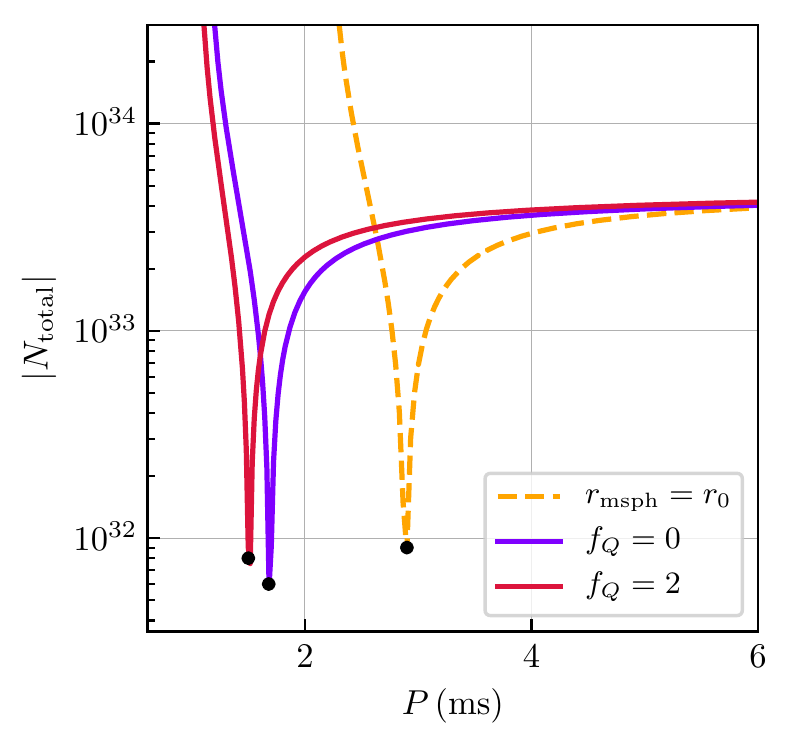}
\caption{Total torque vs. period of the star. Here, we assumed $B_{\rm dipole}=10^{10}\,\mathrm{G}$, $M=1.4\,M_{\sun}$, $R=10\,\mathrm{km}$,
$\alpha=0.1$, $h/r=0.1$, $\dot{M}=5\times 10^{16}\,\mathrm{g/s}$. Accordingly, $\beta=0.05$ when $f_Q=2$. The dashed line represents where
the magnetospheric radius of the star is estimated by the conventional expression given in equation~\eqref{eq:r0} for a pure dipole stellar field.
In the solid lines, the magnetospheric radius is estimated by equation~\eqref{eq:rmsph} for a pure-dipole and quadrudipole stellar fields.
Black dots mark the equilibrium periods.}
\label{fig:torque}
\end{figure}

We report the total torque depending on the period of the star in Fig.~\ref{fig:torque}. 
Even for a pure dipole stellar field, when the magnetospheric radius given in equation~\eqref{eq:rmsph} is employed,
the equilibrium period of the star where the total torque vanishes, alters although the critical fastness parameter remains the same,
because the solution given in equation~\eqref{eq:rmsph} depends on the angular frequency of the star. 
Furthermore, the total torque exerting on the star gets smaller as the quadrupole field's strength increases as observed in \citet{Das22}.

Besides the above torque model, the authors of \citet{Parfrey16,Parfrey17} introduce a torque exerting onto the star due to opening of the magnetic field lines 
which is effective for rapidly rotating stars,
\begin{equation}
N_{\rm wind}=-\zeta^2 \frac{\mu^2}{r_{\rm msph}^2}\frac{\Omega}{c}\,,
\end{equation} 
for the pure dipole stellar field where $\zeta$ parametrises the opening of the magnetic field lines.
Generalising this spin-down torque for quadrudipole fields is beyond the scope of this paper. However,
we can see that this spin-down torque becomes more effective and sustains a smaller equilibrium frequency
when the modification on the magnetospheric radius due to
the spin of the star (see equation~\eqref{eq:rmsph}) is considered
since this modification provides smaller magnetospheric radius,
\begin{equation}
N_{\rm wind}=-\zeta^2 \frac{\mu^2}{r_0^2}\left(1-\frac{\omega_0}{7}\right)^{-4}
\frac{\Omega}{c}
\simeq
-\zeta^2\frac{\mu^2}{r_0^2}\left(1+\frac{4\Omega}{7\Omega_{\rm K}(r_0)}\right)
\frac{\Omega}{c}.
\end{equation}

\section{DISCUSSION} 
\label{sec:discuss}
In the presence of the pure quadrupole stellar field, 
the magnetospheric radius is estimated by using the condition of
the balance of pressures in some works. This condition would provide
the radius where the stellar magnetic field becomes dominant. 
However, using the condition of the balance of stresses realises 
an important difference that the stellar field lines do not hold the 
radial flow of the disc and do not channel the matter onto the poles of the star
when the star has only quadrupole field or
the quadrupole field is dominant over the dipole stellar field around the inner radius of the disc.
Instead, the matter is transferred onto the star through the disc-midplane as 
observed in magnetohydrodynamics simulations \citep{Long07,Long08,Das22}.
We found this limit as $\beta<0.35$.

When the condition of the balance of stresses is used,
the stresses have to be derived from scratch instead of 
generalising the result of the pure dipole stellar field case
by modifying only the radial scaling of the stellar magnetic field.
The dipole and the quadrupole fields produce different magnetic stress terms
since the only non-vanishing component of the quadrupole field at the disc-midplane is
its radial component, unlike the dipole field.
At this point, the generated toroidal field by the quadrupole field is also important.
The toroidal field generated in the presence of the quadrudipole field consists of two terms;
the anti-symmetric part of the toroidal field is generated by the dipole stellar field
while the quadrupole stellar field generates the symmetric part (see equation~\eqref{eq:bphi}).
In the balance of stress equation, the anti-symmetric part of the toroidal field couples with the dipole stellar field
while the symmetric part couples with the quadrupole field (see equations~\eqref{eq:stresses2}-\eqref{eq:bal_main}).
In other words, the contribution of the quadrupole field cannot be realised
unless the toroidal field generated within the disc is calculated.

\begin{figure}
\center
\includegraphics{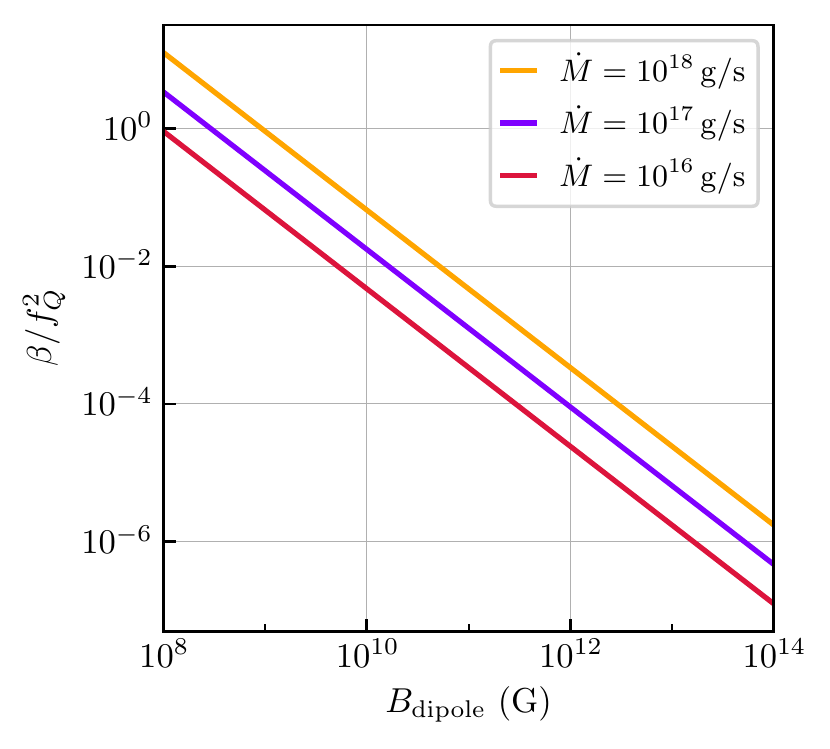}
\caption{$\beta$ (see equation~\ref{eq:beta2}) normalized by the the ratio of the strength of the dipole field to the quadrupole field
vs. the strength of the dipole field for various mass-accretion rates. Here, we assume
$M=1.4\,M_\odot$, $R=10\,\mathrm{km}$, $\alpha=0.1$, $h/r=0.1$.}
\label{fig:beta}
\end{figure}

We report the value of $\beta$ normalized by the $f_Q^2$ depending on the dipole magnetic field strength of the star
for various mass-accretion rates in Fig.~\ref{fig:beta}.
As mentioned above, the presence of the quadrupole field is effective for determining the magnetospheric radius of the disc
and the critical fastness
when $\beta$ is in the range of $\sim 0.1-0.35$.
It is naturally expected that these values are achieved when the dipole magnetic field of the star is weak and
mass-accretion rate is high such that the inner radius of the disc is close to the surface of the star
since the quadrupole field decays more rapidly compared to the dipole field as the radial distance increases.

A quadrupole field which is comparable to the dipole field can be generated
as a result of the mass-accretion onto the star \citep{Suvorov20}.
According to Fig.~\ref{fig:beta}, if $f_Q$ is at the order of unity, the quadrupole field is relevant
for only weakly magnetised stars, i.e.,~$B_{\rm dipole}\sim 10^{8}-10^{10}\,\mathrm{G}$. 
However, limiting the magnetic energy by the gravitational energy constraints 
the value of $f_Q$ to $10^3$ for magnetars and it can be larger 
for smaller dipole magnetic field strengths \citep{Yamasaki22}.
If $f_Q$ is such big, the quadrupole field can be effective 
even for $B_{\rm dipole}\sim 10^{14}\,\mathrm{G}$ as long as the mass-accretion rate is high enough.

Low-mass X-ray binaries (LMXBs) have magnetic fields in a wide range, $B_{\rm dipole}\sim 10^{8}-10^{13}\,\mathrm{G}$ \citep{Revnivtsev15}.
Accordingly, the presence of the quadrupole field might be important in some LMXBs
depending on the value of $f_Q$, the strength of the dipole stellar field and the mass-accretion rate.
On the other hand, pulsating ultra-luminous X-ray sources (PULXs) have stronger magnetic fields, 
$B_{\rm dipole}\gtrsim 10^{12}\,\mathrm{G}$ \citep{Erkut20}. Therefore, the effect of the quadrupole field
on the disc is most likely negligible in PULXs.

In this work, we study the case where the quadrupole field is aligned with the dipole field.
Still, some insight can be gained by comparing magnetic field terms with the aligned case.
For the misaligned case, the radial and poloidal components of the quadrudipole magnetic field become
\begin{align}
B^r_*=&\mu\frac{f_{Q}}{2}\left(3\sin^2(i_Q)-1\right)\frac{R}{r^4}\,,\\
B^\theta_*=&\frac{\mu}{2r^3}\left[1+f_{Q}\sin(2i_Q)\frac{R}{r}\right]\,,
\end{align}
around the disc midplane, i.e. $\theta\rightarrow\pi/2$.
The most apparent difference is 
the presence of quadrupole field in the poloidal component
which makes more difficult to find an analytical expression for the magnetospheric radius.
This contribution is positive when the alignment angle is smaller than $\pi/2$ otherwise it is negative.
On the other hand, the radial component scales with the radial coordinate as the same as the aligned case
but its magnitude varies dependent on the alignment angle.
Its absolute value is smaller than the aligned case when $0<i_Q<3\pi/10$ and $7\pi/10<i_Q<\pi$
while it is bigger when $3\pi/10<i_Q<7\pi/10$. Also, it becomes zero for $i_Q= 0,\pi/5,4\pi/5,\,\pi$.
Accordingly, we can expect that the magnetospheric radius is bigger than the aligned case 
if the alignment angle is smaller than $3\pi/10$.
Moreover, it is even bigger than its value for the pure dipole stellar field when $i_Q=\pi/5$ since the radial component
of the magnetic field vanishes and the poloidal component is bigger.
On the other hand, if the alignment angle is between $\pi/2$ and $7\pi/10$, the magnetospheric radius is smaller than the aligned case.

\section{Conclusion}
We derive the magnetospheric radius for the aligned quadrudipole stellar field as well as 
considering the rotation of the star. The effect of the rotation of the star is
not specific to the quadrupole field, it is relevant even for a pure dipole stellar field. 
Actually, it arises because of the presence of the dipole field.

First, we find the toroidal magnetic field
generated inside the disc from the quadrudipole stellar field. Then, by using 
the generated toroidal field, we write the condition of the balance of stresses in case of the quadrudipole field.
We see that the presence of the quadrupole field decreases the magnetospheric radius due to the direction of the quadrupole field at the disc-midplane.
Furthermore, the disc flow is channelled onto the star through the disc-midplane by following the radial magnetic field
if the quadrupole field is dominant compared to the dipole field. Similarly, the magnetospheric radius decreases
as the spin of the star increases and there is not any solution for the magnetospheric radius beyond the corotation radius.

When the rotational angular frequency of the star or the presence of the quadrupole field
is considered, the magnetospheric radius might reduce to $0.7$ of the conventional magnetospheric 
radius of a pure dipole stellar field given in equation~\eqref{eq:r0}.
The approximate solution given in equation~\eqref{eq:rmsph} is well-consistent with the numerical solution
and the error does not exceed $5\%$ for any values of $\omega_0$ and $\beta$. 

Finally, we calculate the torque exerted onto the star. Since the quadrupole field
generates anti-symmetric toroidal field across the disc midplane, it does not bestow on the magnetic part of the torque.
However, since the presence of the quadrupole field modifies the magnetospheric radius,
the total torque exerted onto the star is different than the pure dipole case.
Depending on the effectiveness of the quadrupole field, the critical fastness parameter might 
shift to $0.995$ from $0.875$. Also, since we consider the effect of the angular frequency difference 
between the star and inner edge of the disc for determining the magnetospheric radius,
the equilibrium period of the star is estimated differently even for the pure dipole stellar field case.

The presence of the quadrupole field might be effective for determining the magnetospheric radius
in LMXBs depending on the value of $f_Q$, the strength of the dipole stellar field and mass-accretion rate.
On the other hand, the effect of the angular velocity of the star is relevant for the systems which are close to the spin equilibrium,
since it is effective when the magnetospheric radius is close to the corotation radius. Both effects might be relevant
for accreting millisecond X-ray pulsars (AMXPs) since these objects are usually in the spin equilibrium and have weak magnetic fields.

In this paper, we consider the quadrupole stellar magnetic field perfectly penetrates 
into the disc. However, it is most likely that there will be a screening factor arising due to
the turbulent motion within the disc. This can be better understood by performing 
the general relativistic magnetohydrodynamics simulations which we plan to
investigate it in a sequence work of \citet{Cikintoglu22}. The same form of the solution 
found in this paper for the magnetospheric radius can still be used even in the presence of such effects
by redefining appropriately $\xi$ and $\beta$ as long as the effect does not vary radially.
\section*{Acknowledgements}
The author thanks K.~Yavuz Ek\c{s}i for his useful comments.
SC acknowledges support from Scientific and Technological Research Council of Turkey (TÜBİTAK) with project number 122F456.
\section*{Data availability}

This is a theoretical paper that does not involve any new data. The
model data presented in this article are all reproducible.

\bibliographystyle{mnras}
\bibliography{refs.bib} 

\begin{thebibliography}{}
\makeatletter
\relax
\def\mn@urlcharsother{\let\do\@makeother \do\$\do\&\do\#\do\^\do\_\do\%\do\~}
\def\mn@doi{\begingroup\mn@urlcharsother \@ifnextchar [ {\mn@doi@}
  {\mn@doi@[]}}
\def\mn@doi@[#1]#2{\def\@tempa{#1}\ifx\@tempa\@empty \href
  {http://dx.doi.org/#2} {doi:#2}\else \href {http://dx.doi.org/#2} {#1}\fi
  \endgroup}
\def\mn@eprint#1#2{\mn@eprint@#1:#2::\@nil}
\def\mn@eprint@arXiv#1{\href {http://arxiv.org/abs/#1} {{\tt arXiv:#1}}}
\def\mn@eprint@dblp#1{\href {http://dblp.uni-trier.de/rec/bibtex/#1.xml}
  {dblp:#1}}
\def\mn@eprint@#1:#2:#3:#4\@nil{\def\@tempa {#1}\def\@tempb {#2}\def\@tempc
  {#3}\ifx \@tempc \@empty \let \@tempc \@tempb \let \@tempb \@tempa \fi \ifx
  \@tempb \@empty \def\@tempb {arXiv}\fi \@ifundefined
  {mn@eprint@\@tempb}{\@tempb:\@tempc}{\expandafter \expandafter \csname
  mn@eprint@\@tempb\endcsname \expandafter{\@tempc}}}

\bibitem[\protect\citeauthoryear{{Alpar}}{{Alpar}}{2012}]{Alpar12}
{Alpar} M.~A.,  2012, \mn@doi [\mnras] {10.1111/j.1365-2966.2012.21172.x},
  \href {https://ui.adsabs.harvard.edu/abs/2012MNRAS.423.3768A} {423, 3768}

\bibitem[\protect\citeauthoryear{{Ardeljan}, {Bisnovatyi-Kogan}  \&
  {Moiseenko}}{{Ardeljan} et~al.}{2005}]{Ardeljan05}
{Ardeljan} N.~V.,  {Bisnovatyi-Kogan} G.~S.,   {Moiseenko} S.~G.,  2005,
  \mn@doi [\mnras] {10.1111/j.1365-2966.2005.08888.x}, \href
  {https://ui.adsabs.harvard.edu/abs/2005MNRAS.359..333A} {359, 333}

\bibitem[\protect\citeauthoryear{{Bilous} et~al.,}{{Bilous}
  et~al.}{2019}]{Bilous19}
{Bilous} A.~V.,  et~al., 2019, \mn@doi [\apjl] {10.3847/2041-8213/ab53e7},
  \href {https://ui.adsabs.harvard.edu/abs/2019ApJ...887L..23B} {887, L23}

\bibitem[\protect\citeauthoryear{{Brice}, {Zane}, {Turolla}  \& {Wu}}{{Brice}
  et~al.}{2021}]{Brice21}
{Brice} N.,  {Zane} S.,  {Turolla} R.,   {Wu} K.,  2021, \mn@doi [\mnras]
  {10.1093/mnras/stab915}, \href
  {https://ui.adsabs.harvard.edu/abs/2021MNRAS.504..701B} {504, 701}

\bibitem[\protect\citeauthoryear{{Campbell}}{{Campbell}}{1992}]{Campbell92}
{Campbell} C.~G.,  1992, \mn@doi [Geophysical and Astrophysical Fluid Dynamics]
  {10.1080/03091929208228282}, \href
  {https://ui.adsabs.harvard.edu/abs/1992GApFD..63..179C} {63, 179}

\bibitem[\protect\citeauthoryear{{Cheng} \& {Taam}}{{Cheng} \&
  {Taam}}{2003}]{Cheng03}
{Cheng} K.~S.,  {Taam} R.~E.,  2003, \mn@doi [\apj] {10.1086/379009}, \href
  {https://ui.adsabs.harvard.edu/abs/2003ApJ...598.1207C} {598, 1207}

\bibitem[\protect\citeauthoryear{{\c{C}{\i}k{\i}nto{\u{g}}lu}, {Ek{\c{s}}i}  \&
  {Rezzolla}}{{\c{C}{\i}k{\i}nto{\u{g}}lu} et~al.}{2022}]{Cikintoglu22}
{\c{C}{\i}k{\i}nto{\u{g}}lu} S.,  {Ek{\c{s}}i} K.~Y.,   {Rezzolla} L.,  2022,
  \mn@doi [\mnras] {10.1093/mnras/stac2510}, \href
  {https://ui.adsabs.harvard.edu/abs/2022MNRAS.517.3212C} {517, 3212}

\bibitem[\protect\citeauthoryear{{Dall'Osso}, {Perna}, {Papitto}, {Bozzo}  \&
  {Stella}}{{Dall'Osso} et~al.}{2016}]{DallOsso16}
{Dall'Osso} S.,  {Perna} R.,  {Papitto} A.,  {Bozzo} E.,   {Stella} L.,  2016,
  \mn@doi [\mnras] {10.1093/mnras/stw110}, \href
  {https://ui.adsabs.harvard.edu/abs/2016MNRAS.457.3076D} {457, 3076}

\bibitem[\protect\citeauthoryear{{Das}, {Porth}  \& {Watts}}{{Das}
  et~al.}{2022}]{Das22}
{Das} P.,  {Porth} O.,   {Watts} A.~L.,  2022, \mn@doi [\mnras]
  {10.1093/mnras/stac1817}, \href
  {https://ui.adsabs.harvard.edu/abs/2022MNRAS.515.3144D} {515, 3144}

\bibitem[\protect\citeauthoryear{{Eksi}, {Andac}, {Cikintoglu}, {Gencali},
  {Gungor}  \& {Oztekin}}{{Eksi} et~al.}{2015}]{Eksi15}
{Eksi} K.~Y.,  {Andac} I.~C.,  {Cikintoglu} S.,  {Gencali} A.~A.,  {Gungor} C.,
    {Oztekin} F.,  2015, \mn@doi [\mnras] {10.1093/mnrasl/slu199}, \href
  {https://ui.adsabs.harvard.edu/abs/2015MNRAS.448L..40E} {448, L40}

\bibitem[\protect\citeauthoryear{{Erkut} \& {Alpar}}{{Erkut} \&
  {Alpar}}{2004}]{Erkut04}
{Erkut} M.~H.,  {Alpar} M.~A.,  2004, \mn@doi [\apj] {10.1086/425169}, \href
  {https://ui.adsabs.harvard.edu/abs/2004ApJ...617..461E} {617, 461}

\bibitem[\protect\citeauthoryear{{Erkut}, {T{\"u}rko{\u{g}}lu}, {Ek{\c{s}}i}
  \& {Alpar}}{{Erkut} et~al.}{2020}]{Erkut20}
{Erkut} M.~H.,  {T{\"u}rko{\u{g}}lu} M.~M.,  {Ek{\c{s}}i} K.~Y.,   {Alpar}
  M.~A.,  2020, \mn@doi [\apj] {10.3847/1538-4357/aba61b}, \href
  {https://ui.adsabs.harvard.edu/abs/2020ApJ...899...97E} {899, 97}

\bibitem[\protect\citeauthoryear{{Ertan}}{{Ertan}}{2017}]{Ertan17}
{Ertan} {\"U}.,  2017, \mn@doi [\mnras] {10.1093/mnras/stw3131}, \href
  {https://ui.adsabs.harvard.edu/abs/2017MNRAS.466..175E} {466, 175}

\bibitem[\protect\citeauthoryear{{Ghosh} \& {Lamb}}{{Ghosh} \&
  {Lamb}}{1978}]{Ghosh78}
{Ghosh} P.,  {Lamb} F.~K.,  1978, \mn@doi [\apjl] {10.1086/182734}, \href
  {http://adsabs.harvard.edu/abs/1978ApJ...223L..83G} {223, L83}

\bibitem[\protect\citeauthoryear{{Ghosh} \& {Lamb}}{{Ghosh} \&
  {Lamb}}{1979a}]{Ghosh79a}
{Ghosh} P.,  {Lamb} F.~K.,  1979a, \mn@doi [\apj] {10.1086/157285}, \href
  {http://adsabs.harvard.edu/abs/1979ApJ...232..259G} {232, 259}

\bibitem[\protect\citeauthoryear{{Ghosh} \& {Lamb}}{{Ghosh} \&
  {Lamb}}{1979b}]{Ghosh79b}
{Ghosh} P.,  {Lamb} F.~K.,  1979b, \mn@doi [\apj] {10.1086/157498}, \href
  {http://adsabs.harvard.edu/abs/1979ApJ...234..296G} {234, 296}

\bibitem[\protect\citeauthoryear{{Ghosh}, {Lamb}  \& {Pethick}}{{Ghosh}
  et~al.}{1977}]{Ghosh77}
{Ghosh} P.,  {Lamb} F.~K.,   {Pethick} C.~J.,  1977, \mn@doi [\apj]
  {10.1086/155606}, \href {http://adsabs.harvard.edu/abs/1977ApJ...217..578G}
  {217, 578}

\bibitem[\protect\citeauthoryear{{Grindlay}, {Camilo}, {Heinke}, {Edmonds},
  {Cohn}  \& {Lugger}}{{Grindlay} et~al.}{2002}]{Grindlay02}
{Grindlay} J.~E.,  {Camilo} F.,  {Heinke} C.~O.,  {Edmonds} P.~D.,  {Cohn} H.,
   {Lugger} P.,  2002, \mn@doi [\apj] {10.1086/344150}, \href
  {https://ui.adsabs.harvard.edu/abs/2002ApJ...581..470G} {581, 470}

\bibitem[\protect\citeauthoryear{{G{\"u}ver}, {G{\"o}{\v{g}}{\"u}{\c{s}}}  \&
  {{\"O}zel}}{{G{\"u}ver} et~al.}{2011}]{Guver11}
{G{\"u}ver} T.,  {G{\"o}{\v{g}}{\"u}{\c{s}}} E.,   {{\"O}zel} F.,  2011,
  \mn@doi [\mnras] {10.1111/j.1365-2966.2011.19677.x}, \href
  {https://ui.adsabs.harvard.edu/abs/2011MNRAS.418.2773G} {418, 2773}

\bibitem[\protect\citeauthoryear{{Illarionov} \& {Sunyaev}}{{Illarionov} \&
  {Sunyaev}}{1975}]{Illarionov75}
{Illarionov} A.~F.,  {Sunyaev} R.~A.,  1975, \aap, \href
  {https://ui.adsabs.harvard.edu/abs/1975A&A....39..185I} {39, 185}

\bibitem[\protect\citeauthoryear{{Kaburaki}}{{Kaburaki}}{1986}]{Kaburaki86}
{Kaburaki} O.,  1986, \mn@doi [\mnras] {10.1093/mnras/220.2.321}, \href
  {https://ui.adsabs.harvard.edu/abs/1986MNRAS.220..321K} {220, 321}

\bibitem[\protect\citeauthoryear{{Kalapotharakos}, {Wadiasingh}, {Harding}  \&
  {Kazanas}}{{Kalapotharakos} et~al.}{2021}]{Kalapotharakos21}
{Kalapotharakos} C.,  {Wadiasingh} Z.,  {Harding} A.~K.,   {Kazanas} D.,  2021,
  \mn@doi [\apj] {10.3847/1538-4357/abcec0}, \href
  {https://ui.adsabs.harvard.edu/abs/2021ApJ...907...63K} {907, 63}

\bibitem[\protect\citeauthoryear{{Krolik}}{{Krolik}}{1991}]{Krolik91}
{Krolik} J.~H.,  1991, \mn@doi [\apjl] {10.1086/186053}, \href
  {https://ui.adsabs.harvard.edu/abs/1991ApJ...373L..69K} {373, L69}

\bibitem[\protect\citeauthoryear{{Li} \& {Wang}}{{Li} \& {Wang}}{1996}]{Li96}
{Li} X.~D.,  {Wang} Z.~R.,  1996, \aap, \href
  {https://ui.adsabs.harvard.edu/abs/1996A&A...307L...5L} {307, L5}

\bibitem[\protect\citeauthoryear{{\deLima{Lima}{De}{de} Lima}, {Coelho},
  {Pereira}, {Rodrigues}  \& {Rueda}}{{\deLima{Lima}{De}{de} Lima}
  et~al.}{2020}]{Lima20}
{\deLima{Lima}{De}{de} Lima} R. C.~R.,  {Coelho} J.~G.,  {Pereira} J.~P.,
  {Rodrigues} C.~V.,   {Rueda} J.~A.,  2020, \mn@doi [\apj]
  {10.3847/1538-4357/ab65f4}, \href
  {https://ui.adsabs.harvard.edu/abs/2020ApJ...889..165D} {889, 165}

\bibitem[\protect\citeauthoryear{{Lipunov}}{{Lipunov}}{1978}]{Lipunov78}
{Lipunov} V.~M.,  1978, \sovast, \href
  {https://ui.adsabs.harvard.edu/abs/1978SvA....22..702L} {22, 702}

\bibitem[\protect\citeauthoryear{{Long}, {Romanova}  \& {Lovelace}}{{Long}
  et~al.}{2007}]{Long07}
{Long} M.,  {Romanova} M.~M.,   {Lovelace} R.~V.~E.,  2007, \mn@doi [\mnras]
  {10.1111/j.1365-2966.2006.11192.x}, \href
  {https://ui.adsabs.harvard.edu/abs/2007MNRAS.374..436L} {374, 436}

\bibitem[\protect\citeauthoryear{{Long}, {Romanova}  \& {Lovelace}}{{Long}
  et~al.}{2008}]{Long08}
{Long} M.,  {Romanova} M.~M.,   {Lovelace} R.~V.~E.,  2008, \mn@doi [\mnras]
  {10.1111/j.1365-2966.2008.13124.x}, \href
  {https://ui.adsabs.harvard.edu/abs/2008MNRAS.386.1274L} {386, 1274}

\bibitem[\protect\citeauthoryear{{Long}, {Romanova}  \& {Lamb}}{{Long}
  et~al.}{2012}]{Long12}
{Long} M.,  {Romanova} M.~M.,   {Lamb} F.~K.,  2012, \mn@doi [\na]
  {10.1016/j.newast.2011.08.001}, \href
  {https://ui.adsabs.harvard.edu/abs/2012NewA...17..232L} {17, 232}

\bibitem[\protect\citeauthoryear{{Miller} et~al.,}{{Miller}
  et~al.}{2019}]{Miller19}
{Miller} M.~C.,  et~al., 2019, \mn@doi [\apjl] {10.3847/2041-8213/ab50c5},
  \href {https://ui.adsabs.harvard.edu/abs/2019ApJ...887L..24M} {887, L24}

\bibitem[\protect\citeauthoryear{{Parfrey}, {Spitkovsky}  \&
  {Beloborodov}}{{Parfrey} et~al.}{2016}]{Parfrey16}
{Parfrey} K.,  {Spitkovsky} A.,   {Beloborodov} A.~M.,  2016, \mn@doi [\apj]
  {10.3847/0004-637X/822/1/33}, \href
  {https://ui.adsabs.harvard.edu/abs/2016ApJ...822...33P} {822, 33}

\bibitem[\protect\citeauthoryear{{Parfrey}, {Spitkovsky}  \&
  {Beloborodov}}{{Parfrey} et~al.}{2017}]{Parfrey17}
{Parfrey} K.,  {Spitkovsky} A.,   {Beloborodov} A.~M.,  2017, \mn@doi [\mnras]
  {10.1093/mnras/stx950}, \href
  {https://ui.adsabs.harvard.edu/abs/2017MNRAS.469.3656P} {469, 3656}

\bibitem[\protect\citeauthoryear{{Pringle} \& {Rees}}{{Pringle} \&
  {Rees}}{1972}]{Pringle72}
{Pringle} J.~E.,  {Rees} M.~J.,  1972, \aap, \href
  {http://adsabs.harvard.edu/abs/1972A%26A....21....1P} {21, 1}

\bibitem[\protect\citeauthoryear{{Psaltis} \& {Chakrabarty}}{{Psaltis} \&
  {Chakrabarty}}{1999}]{Psaltis99}
{Psaltis} D.,  {Chakrabarty} D.,  1999, \mn@doi [\apj] {10.1086/307525}, \href
  {https://ui.adsabs.harvard.edu/abs/1999ApJ...521..332P} {521, 332}

\bibitem[\protect\citeauthoryear{{Rappaport}, {Fregeau}  \&
  {Spruit}}{{Rappaport} et~al.}{2004}]{Rappaport04}
{Rappaport} S.~A.,  {Fregeau} J.~M.,   {Spruit} H.,  2004, \mn@doi [\apj]
  {10.1086/382863}, \href
  {https://ui.adsabs.harvard.edu/abs/2004ApJ...606..436R} {606, 436}

\bibitem[\protect\citeauthoryear{{Revnivtsev} \& {Mereghetti}}{{Revnivtsev} \&
  {Mereghetti}}{2015}]{Revnivtsev15}
{Revnivtsev} M.,  {Mereghetti} S.,  2015, \mn@doi [\ssr]
  {10.1007/s11214-014-0123-x}, \href
  {https://ui.adsabs.harvard.edu/abs/2015SSRv..191..293R} {191, 293}

\bibitem[\protect\citeauthoryear{{Riley} et~al.,}{{Riley}
  et~al.}{2019}]{Riley19}
{Riley} T.~E.,  et~al., 2019, \mn@doi [\apjl] {10.3847/2041-8213/ab481c}, \href
  {https://ui.adsabs.harvard.edu/abs/2019ApJ...887L..21R} {887, L21}

\bibitem[\protect\citeauthoryear{{Romanova}, {Ustyugova}, {Koldoba}  \&
  {Lovelace}}{{Romanova} et~al.}{2002}]{Romanova02}
{Romanova} M.~M.,  {Ustyugova} G.~V.,  {Koldoba} A.~V.,   {Lovelace} R.~V.~E.,
  2002, \mn@doi [\apj] {10.1086/342464}, \href
  {https://ui.adsabs.harvard.edu/abs/2002ApJ...578..420R} {578, 420}

\bibitem[\protect\citeauthoryear{{Romanova}, {Toropina}, {Toropin}  \&
  {Lovelace}}{{Romanova} et~al.}{2003}]{Romanova03}
{Romanova} M.~M.,  {Toropina} O.~D.,  {Toropin} Y.~M.,   {Lovelace} R.~V.~E.,
  2003, \mn@doi [\apj] {10.1086/373990}, \href
  {https://ui.adsabs.harvard.edu/abs/2003ApJ...588..400R} {588, 400}

\bibitem[\protect\citeauthoryear{{Romanova}, {Ustyugova}, {Koldoba}  \&
  {Lovelace}}{{Romanova} et~al.}{2004}]{Romanova04}
{Romanova} M.~M.,  {Ustyugova} G.~V.,  {Koldoba} A.~V.,   {Lovelace} R.~V.~E.,
  2004, \mn@doi [\apjl] {10.1086/426586}, \href
  {https://ui.adsabs.harvard.edu/abs/2004ApJ...616L.151R} {616, L151}

\bibitem[\protect\citeauthoryear{{Scharlemann}}{{Scharlemann}}{1978}]{Scharlemann78}
{Scharlemann} E.~T.,  1978, \mn@doi [\apj] {10.1086/155823}, \href
  {https://ui.adsabs.harvard.edu/abs/1978ApJ...219..617S} {219, 617}

\bibitem[\protect\citeauthoryear{{Shakura} \& {Sunyaev}}{{Shakura} \&
  {Sunyaev}}{1973}]{Shakura73}
{Shakura} N.~I.,  {Sunyaev} R.~A.,  1973, \aap, \href
  {https://ui.adsabs.harvard.edu/abs/1973A&A....24..337S} {500, 33}

\bibitem[\protect\citeauthoryear{{Suvorov} \& {Melatos}}{{Suvorov} \&
  {Melatos}}{2020}]{Suvorov20}
{Suvorov} A.~G.,  {Melatos} A.,  2020, \mn@doi [\mnras]
  {10.1093/mnras/staa3132}, \href
  {https://ui.adsabs.harvard.edu/abs/2020MNRAS.499.3243S} {499, 3243}

\bibitem[\protect\citeauthoryear{{Wang}}{{Wang}}{1987}]{Wang87}
{Wang} Y.~M.,  1987, \aap, \href
  {https://ui.adsabs.harvard.edu/abs/1987A&A...183..257W} {183, 257}

\bibitem[\protect\citeauthoryear{{Wang}}{{Wang}}{1995}]{Wang95}
{Wang} Y.~M.,  1995, \mn@doi [\apjl] {10.1086/309649}, \href
  {https://ui.adsabs.harvard.edu/abs/1995ApJ...449L.153W} {449, L153}

\bibitem[\protect\citeauthoryear{{Wang}}{{Wang}}{1996}]{Wang96}
{Wang} Y.~M.,  1996, \mn@doi [\apjl] {10.1086/310150}, \href
  {https://ui.adsabs.harvard.edu/abs/1996ApJ...465L.111W} {465, L111}

\bibitem[\protect\citeauthoryear{{Yamasaki}, {Ek{\c{s}}i}  \&
  {G{\"o}{\u{g}}{\"u}{\c{s}}}}{{Yamasaki} et~al.}{2022}]{Yamasaki22}
{Yamasaki} S.,  {Ek{\c{s}}i} K.~Y.,   {G{\"o}{\u{g}}{\"u}{\c{s}}} E.,  2022,
  \mn@doi [\mnras] {10.1093/mnras/stac699}, \href
  {https://ui.adsabs.harvard.edu/abs/2022MNRAS.512.3189Y} {512, 3189}

\bibitem[\protect\citeauthoryear{{Yi}}{{Yi}}{1995}]{Yi95}
{Yi} I.,  1995, \mn@doi [\apj] {10.1086/175482}, \href
  {https://ui.adsabs.harvard.edu/abs/1995ApJ...442..768Y} {442, 768}

\makeatother
\end{thebibliography}



\bsp	
\label{lastpage}
\end{document}